\documentclass[twoside,twocolumn,english,9pt]{revtex4-1}
\usepackage[T1]{fontenc}
\usepackage[latin9]{inputenc}
\usepackage[a4paper]{geometry}
\usepackage{textcomp}
\usepackage{amsmath}
\usepackage{amssymb}
\usepackage{graphicx}
\usepackage{esint}

\makeatletter

\usepackage{babel}

\makeatother

\usepackage{babel}
\begin{document}

\title[]{Encoding curved spacetime into initial condition}

\author{Giuseppe Di Molfetta}
\email{giuseppe.dimolfetta@lif.univ-amu.fr}
\affiliation{Aix-Marseille Univ., CNRS, Laboratoire d'Informatique Fondamentale, {\'E}cole centrale de Marseille, Marseille France and Departamento de F{\'i}sica Te{\'o}rica and IFIC, Universidad de Valencia-CSIC,
Dr. Moliner 50, 46100-Burjassot, Spain}

\date{\today}

\begin{abstract}
We prove that conformal curved spacetime can be encoded into the initial wave function and that curved propagation can be simulated on a two-dimensional regular lattice with a finite set of homogeneous unitary operators. We generalize recent results shown in \cite{sabin2017mapping}, where the author transforms flat-spacetime in curved static spacetime via non-unitary rotations. In particular, encoding the metric in the initial quantum state via unitary local rotations, can naturally be useful for quantum simulators due to the efficiency and simplicity of the implementation scheme. We validate our model proving that the continuous limit converges to the Dirac Eq. in curved spacetime. 
\end{abstract}


\maketitle

\paragraph{Introduction -}

In recent decades, quantum simulation has established itself as an area of Physics that combines conceptual issues - such as the impossibility of cloning quantum states, and the existence of correlations stronger than allowed in the classic domain - with applied questions - such as quantum cryptography and quantum computing. From such interaction comes a more operational way of understanding some aspects of quantum mechanics as a description of Nature. For example, in a seminal article of the quantum computing area \cite{feynman1982simulating}, Richard Feynman suggested that computers, that use quantum logic for information processing, could simulate some quantum systems efficiently, even when this is not possible to computers based on
classical logic. To simulate the dynamics of a quantum system usually means to describe the system in terms of qubits, and its dynamics by a succession of logical gates - which are unitary transformations involving at most two qubits at time. A more natural way of describing quantum systems and their dynamics within a computational perspective is given by Quantum Cellular Automata (QCA). As in the quantum circuit model, quantum cellular automata also describe the system as collection of finite dimensional
systems (qubits, or qudits in general), but the dynamics is autonomous and local. This means the evolution of, say, a qubit in a QCA is fully determined by the surrounding qubits and the local interactions between them. Quantum Walks \cite{grossing1988quantum, Y.AharonovL.Davidovich1993} are the one field sector of a QCA. As models of coherent quantum transport, they are interesting both for fundamental quantum physics and for applications. QWs were first conceived as a natural tool to explore graphs, for example for efficient data searching (see e.g. \cite{magniez2011search}) and in particular, an important field of applications is quantum algorithmics \cite{ambainis2003quantum}. A totally new emergent point of view concerning QWs concerns quantum simulation of gauge fields interaction and standard model phenomenology \cite{PhysRevA.94.012335,genske2013electric,Molfetta2016,di2014quantum} and a broad generic class of spacetime metrics \cite{di2014quantum,arrighi2016quantum,di2013quantum}. However, in all of previous results of relativistic quantum simulation, it was necessary to have an infinite set of local unitaries to drive the walker on curved trajectories \cite{arrighi2016quantum,di2013quantum}. Each vertex of the lattice carries the information about the curved spacetime. In this letter, initially inspired by \cite{sabin2017mapping}, we show that, for conformal metrics, this is achievable with just one local encoding of the initial state and local decoding of the final state. All unitary operators between the initial and the final encodings are homogeneous in spacetime. This "homogenization" technique has a clear practical interest because require to use the same quantum simulators, which are already able to reproduce a free massless (1+1) Dirac Eq. \cite{gerritsma2010quantum, lamata2007dirac}. 
In the following, as exemple of interest, we show that any solution of a free massless Dirac Eq. in (1+1)-dimensions can be transformed into a solution of the same equation in any curved spacetime. 

\paragraph{The problem -}

We start here to review, rapidly, how we can model fermion propagation in a (1+1) flat spacetime. Consider a QW defined over discrete time and discrete one dimensional space, labeled respectively by $t\in\mathbb{N}$ and $x\in\mathbb{Z}$. The evolution Eq. reads: 
\begin{equation}
\Psi(t+\Delta t,x) = SQ_{\varepsilon} \Psi(t,x),
\label{eq:defwalkdiscr}
\end{equation}
where
\begin{equation}
Q_{\varepsilon}=\begin{pmatrix}\cos(\varepsilon\theta) & i\sin(\varepsilon\theta)\\
i\sin(\varepsilon\theta) & \cos(\varepsilon\theta),
\end{pmatrix}
\end{equation}
is the quantum coin acting on each internal state of the walker depending
on dimensionless real parameters $(\varepsilon,\theta)$. The operator $S$ is the usual spin-dependent shift operator, acting on each internal state component, $\Psi(t,x)=\{\psi(t,x)^{\uparrow},\psi(t,x)^{\downarrow}\}\in\mathcal{H}_{{spin}_{\Psi}},\otimes\mathbb{Z}$
and defined as follows: 
\begin{equation}
S\Psi(t,x)=\left(\psi^{\uparrow}(t,x+\Delta x),\psi^{\downarrow}(t,x-\Delta x)\right)^{\top}.
\end{equation}
Eq. (\ref{eq:defwalkdiscr}) describes the evolution of a two-level system, and it is well known that Taylor expanding Eq. (\ref{eq:defwalkdiscr}) around $\varepsilon = 0$, we recover the Dirac Eq. in flat spacetime $\eta_{\mu\nu}$, at order $O(\varepsilon)$:
\begin{equation}
i \partial_t \Psi = (i \sigma_z  \partial_x - \theta \sigma_x) \Psi 
\label{eq:Dirac}
\end{equation}
where the parameter $\theta$ corresponds to the Dirac mass ($\hbar = c=1$). When the spacetime is no longer flat, the (1+1)-Dirac Eq. in Eq.(\ref{eq:Dirac}) can be always recast as follows:
\begin{equation}
i (\partial_t + \frac{\dot\Omega}{2\Omega}) \Psi = \left(i \sigma_z (\partial_x + \frac{\Omega^{'}}{2 \Omega}) + \sigma_z \Omega m\right),
\label{eq:DiracCurved}
\end{equation}
where $\Omega\equiv\Omega(x,t)$ is the conformal factor defined by:
\begin{equation}
g_{\mu \nu} = \Omega(t,x) \eta_{\mu \nu}.
\end{equation}
The latter equation holds for any spacetime in (1+1) dimensions and for all conformal spacetime in any spacetime dimensions. In particular, as it has been shown, e.g., in \cite{koke2016dirac}, although in (1+1), this spacetime is not flat, in fact  the Christoffel symbols reads: $\Gamma^{0}_{00} = \Gamma^{0}_{11} = \Gamma^{1}_{01} =\Gamma^{1}_{10} = \dot\Omega\ /\Omega$ and $\Gamma^{0}_{01} = \Gamma^{0}_{10} = \Gamma^{1}_{00} = \Gamma^{1}_{11} = \Omega^{'}\ /\Omega$. Thus, the Ricci curvature does not vanish and reads:
\begin{equation} 
R = g^{\mu\nu}R_{\mu\nu} = 2 ((\dot\Omega/\Omega)^2-\ddot\Omega/\Omega)/\Omega^2.
\end{equation} 
\
Now, let us suppose that we want to encode our metric coefficient, $\Omega(t,x)$, in the wave function $\Psi(t,x)$. This means to compute $\tilde\Psi(t,x) = U_\Omega(t,x) \Psi(t,x)$, where $U_\Omega(t,x)$ is any unitary local operation depending on $\Omega(,x)$, and let us write the Eq. $(\ref{eq:defwalkdiscr})$ for $\tilde\Psi(t,x)$. A simple and straightforward calculation show that this encoding cannot lead to a different dynamical behavior for the probability density of $\tilde \Psi$, because $|\Psi(t,x)|^2 =  |\tilde\Psi(t,x)|^2$, due to the unitarity of  $U_\Omega(t,x)$.

\begin{figure}
\includegraphics[width=1\columnwidth]{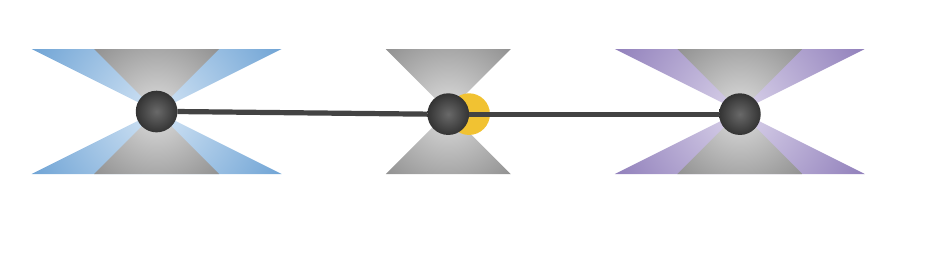}

\caption{(Color online) The coin at each vertex of the lattice carries the local information about the spacetime metric.}
\label{fig:standard} 
\end{figure}

\paragraph{The solution -}

The solution consists in (i) relaxing the unitarity condition performing an isometry on the initial state, which implies doubling the Hilbert space; (ii) encoding the metric and then (iii) come back to the original space by projection onto the initial basis. Let us start with defining a new Hilber space $\mathcal{H}_{{spin}_{\Psi}}\otimes\mathcal{H}_{{spin}_{\Phi}}$, which transcribes in preparing not one but two different particles at the initial state, $\Lambda(t,x) = (\Psi(t,x),\Phi(t,x))$:

\begin{equation}
\Lambda(t+\Delta t,x) =(\mathbb{I}_2\otimes S)\mathcal{Q}_{\varepsilon} \Lambda(t,x).\ 
\label{eq:defwalkdiscrII}
\end{equation}
where 
\begin{equation}
(\mathbb{I}_2\otimes S)\mathcal{Q}_{\varepsilon} = \begin{pmatrix} SQ_{\varepsilon}^{\Psi} & 0 \\ 0 & SQ_{\varepsilon}^{\Phi}    \end{pmatrix}
\end{equation}

Each of this field obeys to a massive Dirac Eq. in flat spacetime, as we recalled in the previous section. An encoding unitary and local operator $U(t,x)$ acting on $\Lambda(t,x)$, such that $\tilde \Lambda$ = $U \Lambda$, may have the form:

\begin{equation}
U = \begin{pmatrix}
\sqrt{(1-\varepsilon^\eta)} N & \sqrt{\varepsilon^\eta} H \\
-\sqrt{\varepsilon^\eta}  T & \sqrt{(1-\varepsilon^\eta)}V
\end{pmatrix},
\end{equation}

where  the necessary conditions must be satisfied:

\begin{align}
(1-\varepsilon^\eta) N^\dagger N + \varepsilon^\eta T^\dagger T= \mathbb{I} \nonumber \\
(1-\varepsilon^\eta) V^\dagger V + \varepsilon^\eta H^\dagger H = \mathbb{I}\nonumber \\
(N^\dagger H - T^\dagger V) = 0,
\label{eq:cond}
\end{align}

and Eqs. (\ref{eq:defwalkdiscrII}) now transcribe in:
\begin{align}
\tilde \Lambda(t+\Delta t,x)= U(t+\Delta t,x)(\mathbb{I}_2\otimes S)\mathcal{Q}_{\varepsilon}U^{\dagger}(t,x)\tilde\Lambda(t,x).
\label{eq:walkwithtilde}
\end{align}

Notice that the necessary conditions (\ref{eq:cond}) guarantee the unitarity just upon $U$. Each 2-dimensional block can be in general non-unitary and, in particular, the parameter $\varepsilon$ allows us to relax the unitarity condition only when $\varepsilon \neq 0$. \\
More explicitly Eqs. (\ref{eq:walkwithtilde}) read:

\begin{align}
\sqrt{1-\varepsilon^\eta} (N^\dagger \tilde\Psi)(t+\Delta t,x) - \sqrt{\varepsilon^\eta} (T^\dagger \tilde\Phi)(t+\Delta t,x)  = \nonumber  \\
 = S \left( Q_{\varepsilon}^{\Psi}  \sqrt{1-\varepsilon^\eta} (N^\dagger \tilde\Psi)(t,x) - Q_{\varepsilon}^{\Phi} \sqrt{\varepsilon^\eta} (T^\dagger \tilde\Phi)(t,x)\right)  \nonumber  \\
\sqrt{\varepsilon^\eta} (H^\dagger \tilde\Phi)(t+ \Delta t,x) + \sqrt{1-\varepsilon^\eta} (V^\dagger \tilde\Psi)(t+\Delta t,x) = \nonumber  \\
 = S \left(Q_{\varepsilon}^{\Phi}  \sqrt{\varepsilon^\eta} (H^\dagger \tilde\Phi)(t,x)+ Q_{\varepsilon}^{\Psi}  \sqrt{1-\varepsilon^\eta} (V^\dagger \tilde\Psi)(t,x) \right).
 \label{eq:FPDE}
\end{align}

\begin{figure}
\includegraphics[width=1\columnwidth]{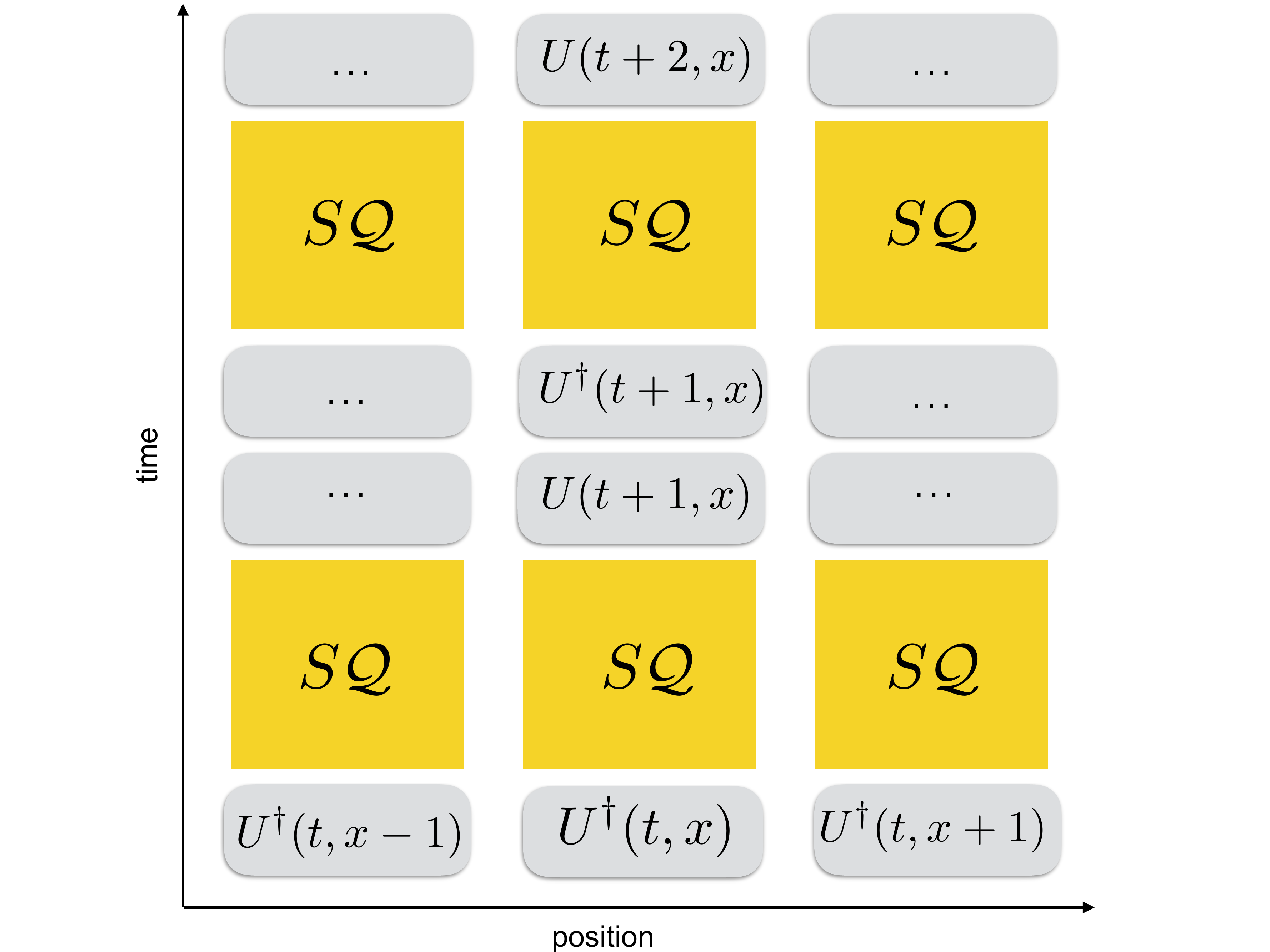}
\caption{(Color online) Quantum Walk on curved spacetime, via encoding and decoding the spacetime. Due to the unitarity of $U$, encoding is needed only at the initial and final time. All the unitary operators (yellow square) between the initial encoding and the final encoding are homogeneous in spacetime}
\label{fig:QW} 
\end{figure}

We want show that these couple of finite difference equation describe curved propagation in each subspace $\mathcal{H}_{{spin}_{\Psi}}$ and $\mathcal{H}_{{spin}_{\Phi}}$. As a validating technique, we compute their continuous limit. We assume that $\tilde{\Psi}$ and $\tilde{\Phi}$ are continuous and at least twice differentiable with respect to both space and time variables, for all sufficiently small values of $\varepsilon$. Proving the existence of the continuous limit implies the following constraint on the coin at the zero order in $\varepsilon$: 
\begin{equation}
\lim_{\varepsilon\rightarrow0}\left[U(\mathbb{I}_2\otimes S)\mathcal{Q}_{\varepsilon}U^{\dagger}\right]=I_{4}.\label{eq:condition}
\end{equation}
In the massless case the Eq. (\ref{eq:condition}) is directly verified because $SQ_{\varepsilon}^{h} = \mathbb{I}_2$, for $h=\Psi,\Phi$, as $\varepsilon\rightarrow0$ and $UU^{\dagger}$= $I_{4}$ by definition.\\

When we Taylor-expand Eqs. (\ref{eq:FPDE}), the leading order at order O($\varepsilon$), transcribes in the following system
of partial differential Eqs. for $\tilde{\Psi}$ and $\tilde{\Phi}$: 
\begin{align}
\tilde\Psi \partial_t N^\dagger + N^\dagger \partial_t \tilde\Psi = \sigma_z (\partial_x N^\dagger) \tilde\Psi +  \sigma_z N^\dagger \partial_x \tilde\Psi  \nonumber \\
\tilde\Phi \partial_t H^\dagger + H^\dagger \partial_t \tilde\Phi = \sigma_z (\partial_x H^\dagger) \tilde\Phi +  \sigma_z H^\dagger \partial_x \tilde\Phi.
 \label{eq: curvedDE}
\end{align}

Now, without loss of generality, we could choose $N^\dagger N$ = $\mathbb{I}_2 \omega^2(t,x)$ for $\tilde\Psi$, where $\omega(t,x)$ is a continuous and differentiable real function and, due to the necessary conditions (\ref{eq:cond}), $H^\dagger H$ = $\mathbb{I}_2 (1-\omega^2(t,x))$ for $\tilde\Phi$.  The Eqs. \label{eq: curvedDE} now coincides with Eq. \ref{eq:DiracCurved}, where $\omega=\sqrt{\Omega}$.
\\
\paragraph{Conclusion -}

We proved that conformal curved spacetime can be encoded into the initial wave functions and that we can simulate curved propagation on a regular lattice with a finite set of homogeneous and stationary unitary operators, between the initial encoding and final decoding. The results obtained here hold for static and dynamical conformal spacetime and in (1+1) hold for any spacetime metrics. The initial and final encodings are unitary, which generalises \cite{sabin2017mapping}, where the author shows how to transform flat-spacetime in curved static spacetime via a non-unitary rotations. All these features make such a model easily realized with all already existing quantum simulation technologies.  
The work presented in this article should naturally be extended in several directions. One should first investigate systematically if this "homogenization" techniques could be extended to non-conformal spacetime \cite{arrighi2016quantum,di2013quantum}. One should also extend the main result of this article to other gauge fields, such as abelian and non-abelian theory. 
As noted earlier, "homogenize" the quantum circuit between the initial encoding and the final encoding, is not just of practical interest, but can have some fundamental issues towards the lattice quantization of any of these gauge fields. Finally, the main result of this article also suggests that concepts from quantum computation and cellular automata may play a key role in understanding of general relativity and its frontier with quantum mechanics. This role should also be investigated thoroughly.

\paragraph{Acknowledgements -}

We benefited from useful discussions with Jose Valle and Armando Perez. In particular we thank Pablo Arrighi for his interest and thorough comments. 

 \bibliographystyle{apsrev4-1}
\bibliography{library}

\end{document}